# Tweets as impact indicators: Examining the implications of automated "bot" accounts on Twitter

Stefanie Haustein[*,1], Timothy D. Bowman[1], Kim Holmberg[2], Andrew Tsou[3], Cassidy R. Sugimoto[3] & Vincent Larivière[1,4]

[*]*stefanie.haustein@umontreal.ca*

[1] École de bibliothéconomie et des sciences de l'information, Université de Montréal, C.P. 6128, Succ. Centre-Ville, Montréal, QC. H3C 3J7 (Canada)

[2] Research Unit for the Sociology of Education, University of Turku, Turku (Finland)

[3] School of Informatics and Computing, Indiana University Bloomington
1320 E. 10th St. Bloomington, IN 47401 (USA)

[4] Observatoire des Sciences et des Technologies (OST), Centre Interuniversitaire de Recherche sur la Science et la Technologie (CIRST), Université du Québec à Montréal, CP 8888, Succ. Centre-Ville, Montréal, QC. H3C 3P8, Canada

**Abstract**
This brief communication presents preliminary findings on automated Twitter accounts distributing links to scientific papers deposited on the preprint repository arXiv. It discusses the implication of the presence of such bots from the perspective of social media metrics (altmetrics), where mentions of scholarly documents on Twitter have been suggested as a means of measuring impact that is both broader and timelier than citations. We present preliminary findings that automated Twitter accounts create a considerable amount of tweets to scientific papers and that they behave differently than common social bots, which has critical implications for the use of raw tweet counts in research evaluation and assessment. We discuss some definitions of Twitter cyborgs and bots in scholarly communication and propose differentiating between different *levels of engagement* from tweeting only bibliographic information to discussing or commenting on the content of a paper.

**Introduction**
The so-called *altmetrics* movement promotes the use of metrics generated from social media platforms as broader and timelier evidence of research impact than citations (Priem, Taraborelli, Groth, & Neylon, 2010). Among these social media platforms, the microblogging platform Twitter has been shown to be one of the most frequently used social media tools that mentions scholarly documents (Costas, Zahedi, & Wouters, 2014; Haustein et al., 2014b; 2014c). As of July 2014, Twitter reported 500 million tweets per day and 255 million monthly active users[1] with 16% of the U.S. population purportedly on Twitter[2]. Scholarly use is proportionally less: it is estimated that between 3% and 15% of researchers utilize this service (Priem & Costello, 2010; Pscheida, Albrecht, Herbst, Minet, & Köhler, 2013; Rowlands, Nicholas, Russell, Canty, & Watkinson, 2011). Therefore, some authors have argued that Twitter is potentially able to capture the impact of scholarly documents on a non-researching public and might, thus, serve as a measure of broader impact of research. However, systematic evidence as to whether tweets are valid markers of actual societal and/or scientific impact is lacking.

Electronic publishing and preprints have accelerated the speed of scholarly communication. Depositing preprints on arXiv has become an integrated part of the publication cycle in Physics, Mathematics, Computer Science, and related fields, providing free and immediate access to research results (Brooks, 2009; Larivière et al., 2014). As tweeting activity related to scholarly documents has been shown to peak and rapidly decline within several days of a document's online availability (Eysenbach, 2011; Shuai, Pepe, & Bollen, 2012), an important share of tweets is assumed to reference the preprint version of a paper, which is available much faster than publication in the journal of record. Comparing Twitter mentions to the two versions of 44,163 arXiv submissions matched to Web of

---

[1] https://about.twitter.com/company
[2] Calculated based on data from:
http://www.sec.gov/Archives/edgar/data/1418091/000119312513400028/d564001ds1a.htm and
http://www.census.gov/population/international/data/

Science (WoS) documents (Haustein et al., 2014a), we discovered the presence of automated Twitter accounts—or "bots"—which distribute links to arXiv preprints automatically.

The presence of bots on Twitter is well-known. Although Twitter has rather strict anti-spam policies, bots continue to flourish in the Twitter environment. This is hardly surprising, given the intense interest in communicating with a large audience as inexpensively as possible. According to an article in *The Japan Times*:

> Twitter-bots are small software programs that are designed to mimic human tweets. Anyone can create bots, though it usually requires programming knowledge. Some bots reply to other users when they detect specific keywords. Others may randomly tweet preset phrases such as proverbs. Or if the bot is designed to emulate a popular person (celebrity, historic icon, anime character etc.) their popular phrases will be tweeted. Not all bots are fully machine-generated, however, and interestingly the term "bot" has also come to refer to Twitter accounts that are simply "fake" accounts. (Akimoto, 2011)

Not all bots are easily identifiable as such, and indeed, social bots are specifically designed to mimic human behavior on social networking sites (Boshmaf, Muslukhov, Beznosov, & Ripeanu, 2011). Twitter bots are not only growing increasingly creative, they are also becoming ubiquitous. It has been found that 16% of Twitter accounts "exhibit a high degree of automation" (Zhang & Paxson, 2011, p. 102), and a study by Chu, Gianvecchio, Wang, and Jajodia (2012) found that 10.5% of Twitter accounts are bots, with an additional 36.2% classified as "cyborgs" (defined as a "bot-assisted human or human-assisted bot"). In addition, an analysis of Russian tweets found that the postings were "structurally dominated by 'instrumental nodes,' accounts being run in the service of marketing" (Kelly et al., 2012, p. 7). Many of these tweets were posted by automated bots, demonstrating that the phenomenon of using bots to spam Twitter with links to questionable sites is not limited to the United States (Kelly et al., 2012).

The majority of Twitter bots follow other tweeters, although this may be due to "follow-spam," which involves following users for the express purpose of spamming them (Mowbray, 2010). An alternate explanation is that this is an attempt by the bots (or, more accurately, their programmers) to "entice the followees themselves to follow the twittering machine back" (Mowbray, 2010, p. 302). Conversely, bots that follow a given tweeter may actually have been commissioned by the tweeter; according to Lotan (2014), the going rate is $5 for 4,000 followers, which makes it relatively inexpensive to amass a large "following."

Many bots contribute positively to Twitter by creating "a large volume of benign tweets, like news and blog updates" (Chu, Gianvecchio, Wang, & Jajodia, 2012, p. 812), and in fact, Twitter itself has been experimenting with automated bots. For example, the @MagicRecs account analyzes its followers' Twitter activities and sends recommendations based on the observed data (Satuluri, 2013). In Japan, earthquake warnings are automatically tweeted (Akimoto, 2011). In the aftermath of a particularly severe earthquake, this automated account proved to be a critical source of information, as many other forms of communication were temporarily unavailable to the general public.

Of course, automated bots are not always used for such benevolent purposes. Apart from general spam (Chu, Gianvecchio, Wang, & Jajodia, 2012), bots have been used to spread viruses via Twitter (e.g., Mirani, 2014), and some automated programs are even capable of mimicking human behavior to such a degree that they can "infiltrate online communities, build up trust over time and then send personalized messages to elicit information, sway opinions and call to action" (Boshmaf, Muslukhov, Beznosov, & Ripeanu, 2012, p. 12). Additionally, the use of Twitter bots is not restricted to advertising agencies or companies wishing to promote their products; in 2011, a contentious election in Russia prompted the use of "thousands of Twitter accounts […] to drown out genuine dissent" (BBC News, 2011, para. 4).

Applications such as the Truthy project's BotOrNot[3] determine the probability of an account being a social bot based on statistical learning (Ferrara, Varol, Davis, Menczer & Flammini, 2014). Other automated methods of analyzing Twitter accounts have (perhaps predictably) found lower incidences of bot posting in accounts that are verified, as well as those that are highly followed. Similarly, automated bots were less likely to tweet about "trending topic search results," presumably because of the ephemeral and unpredictable nature of such phenomena (Zhang & Paxson, 2011, p. 111).

"Robot tweeting" has been listed as a major concern regarding the validity of altmetrics (Darling, Shiffman, Côté, & Drew, 2013) and the ease of gaming social media has been raised as a potential limitation of altmetrics (Cheung, 2013; Priem, 2014). However, with the exception of Shuai, Pepe and Bollen (2012), who report parenthetically that they filtered out 53% of a sample of tweets to arXiv preprints which seem to originate from bot accounts, Twitter bots in scholarly communication have not been the subject of empirical investigations.

In our previous work, we analyzed Twitter activity of arXiv and journal versions of 44,163 scientific papers mentioned in 50,068 tweets (Haustein et al., 2014a). The high percentage of papers with at least one tweet in high energy physics (HEP; 69% to 81%) revealed a series of automatic Twitter accounts such as @hep_th, @hep_ph, @hep_ex, and @hep_lat, which "tweet the new submissions on arXiv…10-60 minutes after the daily update of arXiv/new."[4] Further investigations show that the source code behind these automated accounts was made available on GitHub[5] and has been used to create several other Twitter accounts.[6] As these accounts automatically tweet any new submission without human selection, they undermine the function of tweet counts as a filter or indicator of impact as suggested in the altmetrics manifesto (Priem et al., 2010). Since tweets originate from a machine rather than from the direct action of a human being, the presence of Twitter "bots" brings into question the usefulness of tweets as filters and impact measures. However, a systematic analysis regarding the distinctive types of automated Twitter accounts and the extent of bot accounts in scholarly communication is lacking. This communication discusses preliminary findings regarding the extent of Twitter bot accounts in scholarly communication and implications regarding their effects on tweets as impact indicators.

**Methods**
The Twitter search function via Twitter.com was used to search for accounts containing "arXiv" in the Twitter handle, display name, or account description in May 2014. The result list, which contained 90 accounts, was manually coded by two of the authors. Accounts that contained the keyword but did not refer to the repository (32) or that did not distribute links to arXiv papers (7) were disregarded. The remaining 51 accounts, which tweeted links to papers on arXiv, were classified in one of the following three categories:

1) platform feed: automated feed of papers from an arXiv section or subsection such as @hep_th[7] for HEP - Theory or @arXiv_cs[8] for Computer Science; platform-based feeds tweeting everything published in an arXiv subject area, triggered by arXiv RSS feed[9]

---

[3] http://truthy.indiana.edu/botornot/
[4] http://en.misho-web.com/phys/hep_tools.html#arxiv_speaker
[5] https://github.com/misho104/arxiv_speaker
[6] Except for @hep_th, @hep_ph, @hep_ex and @hep_lat we found 32 Twitter accounts (e.g., @mathAPb, @mathCObot, @mathPRb) for the mathematical subcategories in arXiv, which acknowledge that they modified the script by @misho.
[7] Twitter account description of @hep_th: "arXiv new submissions on hep-th. Feedbacks to @misho are welcome. [README]http://bit.ly/arxiv_twitter [See also] @hep_ex @hep_lat @hep_ph en.misho-web.com/phys/hep_tools…"
[8] Twitter account description of @arXiv_cs: "arXiv cs (computer science) - IFTTT – Twitter Computer export.arxiv.org/rss/cs?mirror=…"
[9] http://arxiv.org/help/rss

2) topic feed: automated feed of papers relevant to a certain topic such as @DMHunters[10] for dark matter or @fly_papers[11] for drosophila research; keyword-based feeds, triggered by keyword specific searches
3) selective/qualitative: some sort of qualitative selection such as @Awesome_Ph[12] or @tweprints[13]; human selection of "interesting" papers

While categories 1) and 2) can be classified as completely automated and non-selective (except for an initial selection of the platform or topic) and thus qualify as Twitter bots, accounts classified in category 3) can be identified as human and selective. The Twitter bots, i.e. platform and topic feeds, made up 9% of all 50,068 tweets to the 2012 arXiv/WoS paper set (Haustein et al., 2014a), even though most of them were created after 2012 (Figure 1), which suggests that the percentage could be higher for more recent papers.

For each of the accounts we identified the number of tweets, number of followers and followees as well as the date of the first[14] and last tweet. The average number of tweets per day was calculated in order to compare the average tweeting activity. Moreover, we obtained the Truthy *Bot or Not?* score, which on a scale from 0% to 100% indicates the probability of a Twitter account to be human or a social bot.[15] *Bot or Not?* takes into account more than 1,000 tweeting characteristics including the appearance of tweets, retweets and mentions, the follower and followee networks as well as tweet content and sentiment (Ferrara et al., 2014).

**Results and Discussion**
Forty-three (84%) of the 51 accounts were identified as platform feeds, while topic feeds and selective/qualitative accounts each appeared 4 times (Table 1). The automated platform and topic feeds produced 87,389 and 10,040 tweets, respectively, which amounts to comparable averages of 2,032 and 2,510 tweets per account, or 4.6 and 7.1 tweets per account per day. As would be expected, the selective/qualitative Twitter accounts tweeted much less than the automated accounts: 770 tweets on average, or 2.2 on average per day. As visualized by the line width in Figure 1 and the size of the data points in Figure 2, the number of tweets per account per day and overall tweets per account range from 0.4 to 21.3 and 5 to more than 10,000, respectively. This variation is also emphasized by high standard deviations (Table 1).

**Table 1** Number of accounts, tweets and tweets per day, mean number of followers, following and *Bot or Not?* score per type of Twitter account. Standard deviations are provided in parentheses.

| type of account | number accounts | sum tweets | mean tweets/day | mean followers | mean following | mean *Bot or Not?* score |
|---|---|---|---|---|---|---|
| platform feed | 43 | 87,389 | 4.6 (4.3) | 34.9 (64.7) | 0.6 (2.8) | 34% (7%) |
| topic feed | 4 | 10,040 | 7.1 (5.6) | 527.0 (662.2) | 491.5 (981.0) | 37% (22%) |
| selective/qualitative | 4 | 3,081 | 2.2 (2.0) | 361.8 (446.2) | 50.5 (101.0) | 35% (17%) |
| *all* | 51 | 100,510 | 4.6 (4.4) | 99.1 (255.9) | 43.0 (275.7) | 34% (10%) |

---

[10] Twitter account description of @DMHunters: "Dark Matter Hunters is a website dedicated to Dark Matter research. We provide a daily digest of Arxiv preprints, among other stuff."
[11] Twitter account description of @fly_papers: "Daily RSS feed for #Drosophila papers in #PubMed and #arXiv. Image of D. hydei from http://en.wikipedia.org/wiki/Drosophila_hydei …"
[12] Twitter account description of @Awesome_Ph: "The most awesome(st) of Astro-ph every day! Few one line summaries of papers on the ArXiV that caught my attention. Brought to you by the Extronomer."
[13] Twitter account description of @tweprints: "Interesting arXiv papers from he Twitter community, UK"
[14] The date of the first tweet was obtained from https://discover.twitter.com/first-tweet
[15] The BotOrNot score was obtained from http://truthy.indiana.edu/botornot

As shown in Figure 1, the first Twitter account related to arXiv tweeted for the first time in June 2009 and was identified as selective/qualitative. Like five other accounts it stopped tweeting after a couple of weeks or months. Of these six inactive accounts, three were coded as selective/qualitative, two as platform feeds and one as topic feed. Thus, in June 2014 only one active selective/qualitative account remained; this account has the second highest number of followers (977).

The majority of platform feeds started sending links to arXiv papers in March and April 2013, had less than 22 followers, 0 followees, and tweeted between 1 and 7 times per day. These accounts are the 32 bots described above, which cover arXiv submissions from the different mathematical arXiv subcategories. They are based on the HEP Twitter bots created in June 2012 and are maintained by a mathematics professor, who lists them as "Twitter arXiv bots" under developments in his CV.[16]

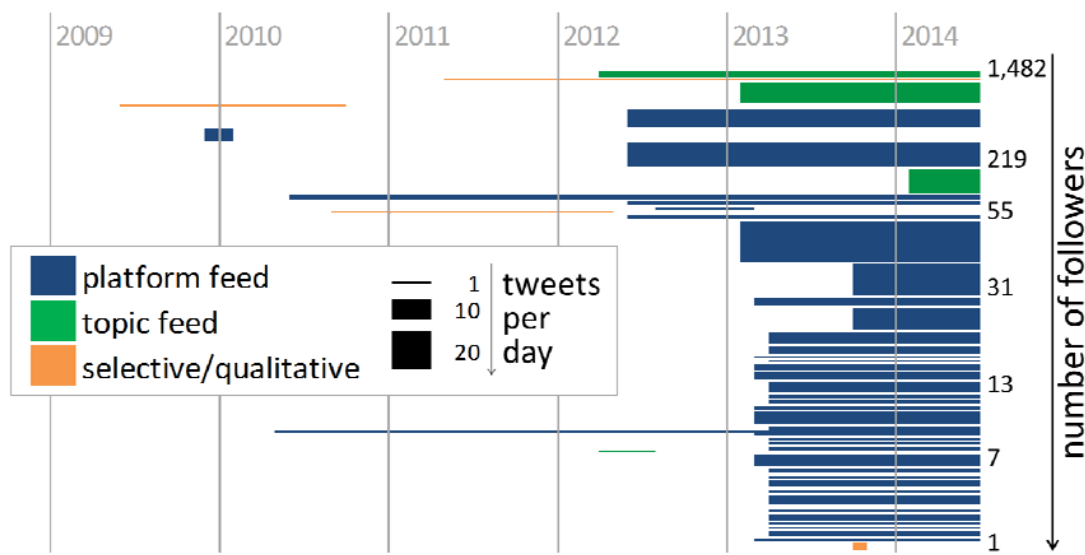

**Figure 1** Timeline of average tweeting activity from first to last tweet for 51 Twitter accounts.

As visualized in Figure 1 and Table 1, topic feeds and selective/qualitative accounts seem to be the most popular among Twitter users, while platform feeds attract as few as 35 followers on average. The latter hardly follow any users (0.6 following on average); 39 accounts did not follow any user, three followed two or three, and one followed 18. That points to the fact that these platform feeds behave differently from normal Twitter bots, which, as described above, often follow other users in order to be followed back or to create "follow-spam" (Mowbray, 2010). The absence of followees of platform feeds also suggests that Twitter users are not interested in or do not know about what is basically a redistribution of the arXiv RSS feed per arXiv subject area or subcategory. It is very likely that these platform feeds have mainly been created because it is easy to do so. Account holders maintain the automatic script but do not directly engage or interact with other users on Twitter. On the other hand, the topic feeds attract on average 491.5 followees, which shows either that Twitter users are actually interested in receiving alerts for papers relevant to certain topics through Twitter or that the topic feeds successfully obtain followers by following them first, which would indicate some engagement on the platform or a behavior observed for other Twitter bots.

Automated Twitter accounts distributing scientific papers without any qualitative selection such as the identified platform feeds and topic feeds undermine the suggested usefulness of tweets as filters of the impact of papers (Priem et al., 2010) and instead merely function as automatic dissemination tools. Thus it would be valuable to identify these bots and distinguish them from accounts maintained by human beings. Since the Truthy *Bot or Not?* tool has been designed to detect social bots, we determined the *Bot or Not?* score for the 51 arXiv Twitter accounts. As shown in Table 1 and Figure

---

[16] http://so.okada.gweb.io

2, the majority of scores were below 50%, indicating that these accounts behaved more like humans than social robots. Only three accounts were considered more likely to be bots than humans. In fact, the highest scores of 64% and 78% were obtained by a topic and a selective/qualitative feed with the highest number of followers, while the accounts with the seven lowest scores (≤26%) were platform feeds. Possible reasons for low scores could be the low number of followers, the absence of followees, complex language of tweet content (i.e., paper titles), etc. This again emphasizes that bots tweeting scientific documents do not behave like social bots, which were the focus of the Truthy *Bot or Not?* project, and are thus not identifiable with the tool.

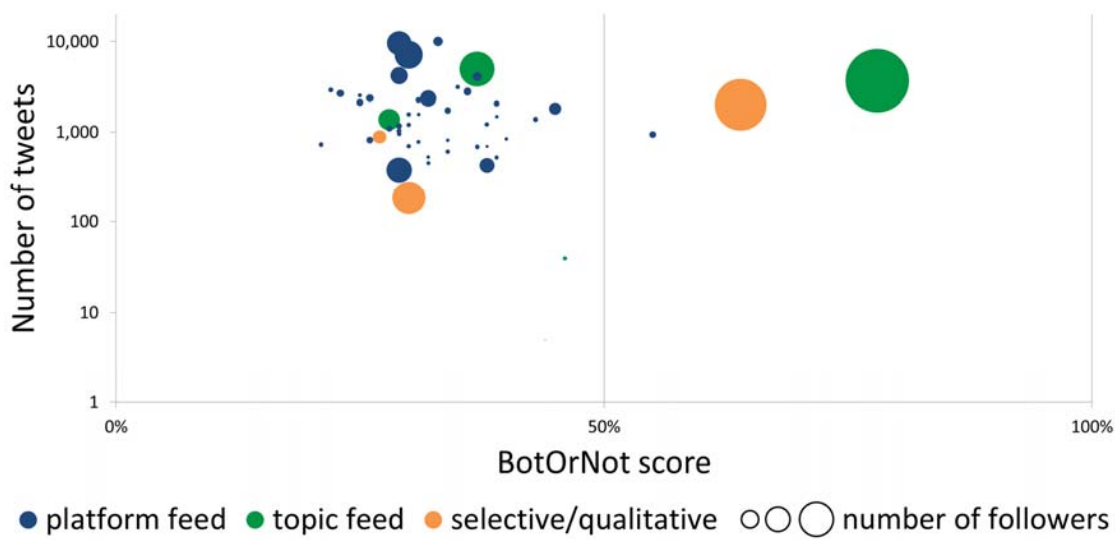

**Figure 2** Number of tweets and followers, *Bot or Not?* score, and type of account for 51 Twitter accounts.

**Conclusion and Future Directions**
The fact that the automated accounts identified in this study at least amount to 9% of tweets to arXiv 2012 submissions published in journals covered by WoS is alarming, particularly as researchers begin putting altmetrics on their CVs (Piwowar & Priem, 2013) and funders start using Twitter impact as indicators of societal impact. In addition, many other automated Twitter accounts cannot be as easily identified through specific keywords as the arXiv Twitter bots examined here. Although accounts that automatically disseminate links to scientific papers behave differently than common social bots and are probably not (yet) created with the intention of gaming the system but as easy to maintain distribution channels, they influence tweet counts. These counts are increasingly being reported and used by scholarly publishers and journals, authors and funders as indicators of impact without reflecting upon or differentiating between the origins of these counts.

Based on exploratory analyses of Twitter accounts disseminating links to scientific papers, we propose to differentiate between human, cyborg, and bot accounts (Figure 3). While human tweeting behavior involves selection and qualitative judgment as to whether to tweet a link to a paper or not (i.e., the selective/qualitative category), a bot does so automatically and non-selectively apart from an initial selection of sources (i.e., the platform and topic feeds). A cyborg (i.e., a computer-assisted human as defined by Chu et al. (2012)) shows a mix of human and bot behavior (i.e., the Twitter account publishes automated as well as tweets created by a person). All three account types can be initiated either by a person (e.g., scientists, students, teachers, practitioners, interested laymen, etc.) or an organization (e.g., scientific societies, publishers, journals, universities, research groups, funding organizations, newspapers, etc.).

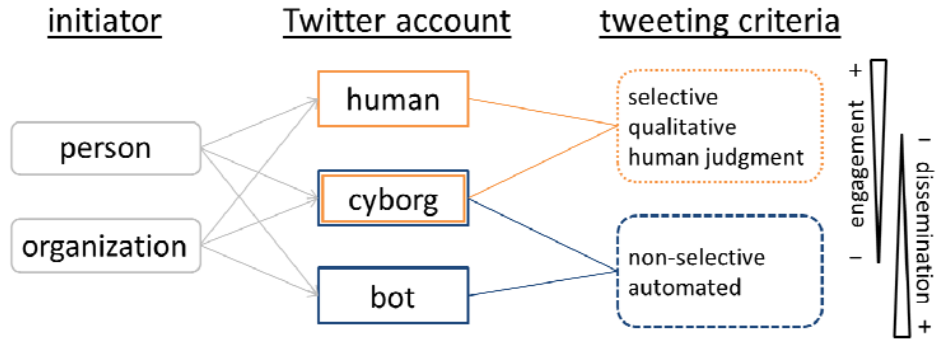

**Figure 3** Differentiation between human, cyborg and bot Twitter accounts in scholarly communication.

Future research should seek to quantify the amount of tweets to scientific articles distributed by human, cyborg and bot accounts. As shown, current automated methods (such as *Bot or Not?*) were insufficient in identifying these accounts. One could imagine an arms race to identify bot accounts as creators of bot programs adapt to avoid detection. According to Campbell's law (Campbell, 1976), misbehavior and corruption will occur, if tweet counts are used as social indicators. However, this arms race should not hinder but actually demands further research to identify bot and human tweets. We suggest incorporating *level of engagement* as a criterion in the identification of automated tweets; namely, the degree to which the content of the tweet demonstrates manipulation beyond copying bibliographic data. For example, Figure 4 displays tweets with high and low levels of engagement, by the similarity between the title of the tweeted paper and the tweet text[17]. Such analyses distinguish between *low* and *high engagement* with the paper (Figure 4). The former seems to be typical of bots, as most of them automatically tweet the paper title but also identifies tweets by regular users when disseminating the bibliographic information without discussing the paper.

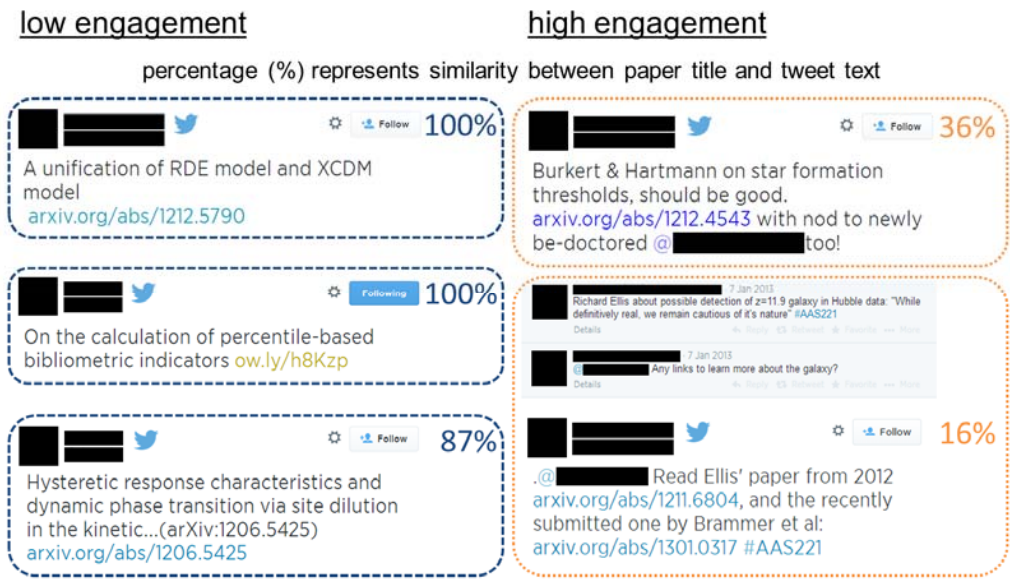

**Figure 4** Differentiation between levels of engagement of tweets based on percentage of similarity between paper title and tweet text.

It could be argued that automated Twitter feeds do not matter when it comes to identifying high impact papers and that a simple solution would be to subtract one tweet from each paper. This would be true if automated accounts were distributed equally. However, the same document can be covered by more than one platform and/or topic feed, and many additional automated accounts by journals, publishers, scientific societies, and authors exist. In an environment where a few tweets boost an article into the 90th percentile of most tweeted papers[18], non-selective automated tweets should at least be identified, if not removed entirely. Moreover, instead of relying on raw counts, tweets could be categorized according to types of Twitter users (including number of followers) and by levels of engagement—from mere dissemination to intense discussion—in order to differentiate by whom and how a paper was tweeted and which audiences it could potentially reach. This differentiated approach would lead to a better understanding of the kind of impact measured through tweets.

---

[17] We used the *similar_text* function in php (http://php.net/manual/en/function.similar-text.php) to compare the paper title from arXiv with the tweet text without URLs, hashtags and usernames (both lowercase).

While this work has focused exclusively on a single social media platform and preprint repository, it should be illustrative of the potential gaming on social media platforms that can systematically bias the evaluation of science. Further research should be conducted on Twitter and other platforms to shed light on the mechanisms underlying social media counts to their meaning and usefulness as impact indicators.

**Acknowledgements**
This work was supported by the Alfred P. Sloan Foundation Grant # G-2014-3-25 "to support greater understanding of social media in scholarly communication and the actual meaning of various altmetrics".

**References**

Akimoto, A. (2011). Japan the Twitter nation. *The Japan Times*, 18. Retrieved from http://www.japantimes.co.jp/life/2011/05/18/digital/japan-the-twitter-nation/#.U7LzA_l5OSo

BBC News (2011, December 9). Russian Twitter political protests 'swamped by spam.' Retrieved from http://www.bbc.com/news/technology-16108876

Boshmaf, Y., Muslukhov, I., Beznosov, K., & Ripeanu, M. (2011). The socialbot network. In *Proceedings of the 27th Annual Computer Security Applications Conference* (p. 93).

Boshmaf, Y., Muslukhov, I., Beznosov, K., & Ripeanu, M. (2012). Key challenges in defending against malicious socialbots [Abstract]. In *Proceedings of the 5th USENIX Conference on Large-scale Exploits and Emergent Threats, LEET'12*, Berkeley, CA (p. 12).

Brooks, T. C. (2009). Organizing a research community with SPIRES: Where repositories, scientists and publishers meet. *Information Services and Use*, 29(2,3), 91–96. doi:10.3233/ISU-2009-0596

Campbell, D. T. (1976). Assessing the impact of planned social change. *Occasional Paper Series*, 8. Hanover, NH: Public Affairs Center, Dartmouth College.Cheung, M. K. (2013). Altmetrics: Too soon for use in assessment. *Nature*, 494(7436), 176. doi:10.1038/494176d

Chu, Z., Gianvecchio, S., Wang, H., & Jajodia, S. (2012). Detecting Automation of Twitter Accounts: Are You a Human, Bot, or Cyborg? *IEEE Transactions on Dependable and Secure Computing*, 9(6), 811-824.

Costas, R., Zahedi, Z., & Wouters, P. (2014). Do "altmetrics" correlate with citations? Extensive comparison of altmetric indicators with citations from a multidisciplinary perspective. *Journal of the Association for Information Science and Technology*. doi: 10.1002/asi.23309

Darling, E. S., Shiffman, D., Côté, I. M., & Drew, J. A. (2013). The role of Twitter in the life cycle of a scientific publication. *Ideas in Ecology and Evolution*, 6(1), 32-43. doi: 10.4033/iee.2013.6.6.f

Eysenbach, G. (2011). Can tweets predict citations? Metrics of social impact based on Twitter and correlation with traditional metrics of scientific impact. *Journal of Medical Internet Research*, 13(4), e123. doi:10.2196/jmir.2012

Ferrara, E., Varol, O., Davis, C., Menczer, F. & Menczer, A. (2014). The rise of social bots. Retrieved from: http://arxiv.org/pdf/1407.5225v1.pdf

Haustein, S., Bowman, T. D., Macaluso, B., Sugimoto, C. R., & Larivière, V. (2014a). Measuring Twitter activity of arXiv e-prints and published papers. In *altmetrics14: expanding impacts and

---

[18] Percentile distribution based on the number of tweets per paper in a certain cohort such as journal, publication age and/or discipline as applied by Altmetric.com and ImpactStory.org.

*metrics. An ACM Web Science Conference 2014 Workshop*. Bloomington, IN. doi:10.6084/m9.figshare.1041514

Haustein, S., Larivière, V., Thelwall, M., Amyot, D., & Peters, I. (2014b). Tweets vs . Mendeley readers: How do these two social media metrics differ ? *it - Information Technology*, 56(5), 207-215. doi: 10.1515/itit-2014-1048

Haustein, S., Peters, I., Sugimoto, C. R., Thelwall, M., & Larivière, V. (2014c). Tweeting biomedicine: an analysis of tweets and citations in the biomedical literature. *Journal of the American Society for Information Science and Technology*, 65(4), 656–669. doi:10.1002/asi.23101

Kelly, J., Barash, V., Alexanyan, K., Etling, B., Faris, R., Gasser, U., & Palfrey, J. (2012). Mapping Russian Twitter. *Berkman Center Research Publication*, (2012-3). Retrieved from: http://cyber.law.harvard.edu/sites/cyber.law.harvard.edu/files/Mapping_Russian_Twitter_2012.pdf

Larivière, V., Sugimoto, C. R., Macaluso, B., Milojević, S., Cronin, B., & Thelwall, M. (2014). arXiv E-prints and the journal of record: An analysis of roles and relationships. *Journal of the Association for Information Science and Technology*, 65(6), 1157–1169. doi:10.1002/asi.23044

Lotan, G. (2014, May 31). Mining Twitter gold, at five bucks a pop. *Los Angeles Times*. Retreived from: http://www.latimes.com/opinion/op-ed/la-oe-0601-lotan-buying-followers-20140601-story.html

Mowbray, M. (2010). The Twittering Machine. In *Proceedings of the 6th International Conference on Web Information Systems and Technologies, WEBIST* (2) (pp. 299-304). Retrieved from: http://shiftleft.com/mirrors/www.hpl.hp.com/techreports/2010/HPL-2010-54.pdf

Piwowar, H., & Priem, J. (2013). The Power of Altmetrics on a CV. *Bulletin of Associatioon for Infomation Science and Technology*, 39(4), 10–13. Retrieved from http://asis.org/Bulletin/Apr-13/AprMay13_Piwowar_Priem.html

Priem, J. (2014). Altmetrics. In *B. Cronin & C. R. Sugimoto (Eds.), Beyond Bibliometrics: Harnessing Multi-dimensional Indicators of Performance* (pp. 263–287). Cambridge, MA: MIT Press.

Priem, J., & Costello, K. L. (2010). How and why scholars cite on Twitter. *Proceedings of the American Society for Information Science and Technology*, 47(1), 1–4. doi:10.1002/meet.14504701201

Priem, J., Taraborelli, D., Groth, P., & Neylon, C. (2010). Alt-metrics: a manifesto. Retrieved from: http://altmetrics.org/manifesto

Pscheida, D., Albrecht, S., Herbst, S., Minet, C., & Köhler, T. (2013). Nutzung von Social Media und onlinebasierten Anwendungen in der Wissenschaft. Erste Ergebnisse des Science 2.0-Survey 2013 des Leibniz-Forschungsverbunds „Science 2.0". Retrieved from http://www.qucosa.de/fileadmin/data/qucosa/documents/13296/Science20_Datenreport_2013_PDF_A.pdf

Rowlands, I., Nicholas, D., Russell, B., Canty, N., & Watkinson, A. (2011). Social media use in the research workflow. *Learned Publishing*, 24(3), 183–195. doi:10.1087/20110306

Satuluri, V. (2013, September 24). Stay in the know [blog post]. Retrieved from https://blog.twitter.com/2013/stay-in-the-know

Shuai, X., Pepe, A., & Bollen, J. (2012). How the scientific community reacts to newly submitted preprints: article downloads, Twitter mentions, and citations. *PLoS ONE*, 7(11), e47523. doi:10.1371/journal.pone.0047523

Thelwall, M., Haustein, S., Larivière, V., & Sugimoto, C. R. (2013). Do altmetrics work? Twitter and ten other candidates. *PLoS ONE*, 8(5), e64841. doi:10.1371/journal.pone.0064841

Zhang, C. M., & Paxson, V. (2011). Detecting and analyzing automated activity on Twitter. In *Proceedings of the 12th International Conference on Passive and Active Measurement PAM'11* (pp. 102-111). Springer Berlin Heidelberg.